\documentclass{nuws}
\usepackage{times}
\usepackage{graphicx} % when using Latex and dvips

\newcommand{\dd}{\ensuremath\mathrm{d}}
\newcommand{\bvec}[1]{\ensuremath\mbox{\boldmath $#1$}}

\begin{document}
\title{Neutrino Flux Bounds and Prospects for High Energy and
Ultrahigh Energy Neutrino Source Detection}
\author[1]{C. Hettlage} \affil[1]{University of G\"ottingen}
\author[2]{K. Mannheim} \affil[2]{University of W\"urzburg}

\correspondence{C. Hettlage (hettlage@uni-sw.gwdg.de)}
\firstpage{1}
\pubyear{2001}

% \titleheight{11cm} % uncomment and adjust in case your title block
                     % does not fit into the default and minimum 7.5 cm

\maketitle

\begin{abstract}
After briefly reviewing various hadronic neutrino source models, we show how
to construct generic upper flux bounds. We then turn to the problem of
neutrino propagation through the inner Earth and neutrino detection in
water-based \v{C}erenkov detectors. Applying the formalism thus developed to
the Mannheim-Protheroe-Rachen (MPR) and the Waxman\,\&\,Bahcall flux bounds, we
find that event rates of several hundred to thousand events per year might
be possible in next-generation neutrino telescopes. However, a tomography of
the inner Earth will face severe constraints due to the statistical error of
the event rates to be expected.
\end{abstract}

\section{Hadronic neutrino source models}
Models for sources producing ultrahigh energy neutrinos can be divided into
hadronic and non-hadronic ones, depending on whether protons play a key part
in the neutrino production. In this work we shall limit ourselves to the
former class of models, referring the reader to \citet{LM00}
for a discussion of non-hadronic source models such as the decay of relic
particles.

Independent of the specific source under consideration, the main idea behind
hadronic neutrino production is that a beam of accelerated protons (taken to
be of power-law form, $n_{p}(E_{p})\propto E_{p}^{-s}$) hitting a
target produces pions, which subsequently decay and thus yield
neutrinos. Here, the target may consist of either protons or photons.

In case of a target made of protons, if the pions decay prior to any
interaction with the surrounding medium, the resulting neutrino production  spectrum is of
the same form as the original spectrum of accelerated protons, i.e. $Q_{\nu}^{pp}(E_{\nu})\propto E_{\nu}^{-s}$. Otherwise, the
neutrino spectrum is steepened \citep{RM98}.

An important example of a photon target is given by the cosmic background
radiation, which leads to the Greisen-Zatsepin-Kuzm\'{\i}n cutoff. In neutrino sources, target
photons may be produced by means of synchrotron radiation. Assuming that the
synchrotron photon spectrum is of the form $N_{\gamma}(E_{\gamma})\propto
E_{\gamma}^{-\alpha}$, one may show that the corresponding neutrino spectrum is also a
power law, $Q_{\nu}^{p\gamma}(E_{\nu})\propto E_{\nu}^{-(s-\alpha)}$ \citep{LM00}.

In the following, we mention several examples of source models which have
been proposed for ultrahigh energy neutrino production:

\subsubsection*{Terrestrial atmosphere:}
Neutrinos produced by cosmic-ray interactions in the terrestrial atmosphere
can be detected with present-day water-based \v{C}erenkov telescopes and thus
serve as an important benchmark \citep{Dom01,And01}.

\subsubsection*{Sun:}
The same processes responsible for neutrino production in the terrestrial
atmosphere can also take place in the Sun. However, the neutrino flux thus to
be expected is so low that it will only be discernible with next-generation
neutrino detectors \citep{MKT91,HML00}.

\subsubsection*{Crab nebula:}
Neutrons resulting from the photodisintegration of iron nuclei may be
accelerated in the outer gap of the Crab pulsar and may subsequently decay into
protons. If these interact with matter in the nebula, pions and thus
neutrinos can be produced \citep{BP97}.

\subsubsection*{Supernova remnants:}
Protons may be shock accelerated and by means of collisions in the remnant may
produce pions and thus neutrinos \citep{GPS98}.

\subsubsection*{AGNs:}
In active galactic nuclei (AGNs), protons may be shock accelerated either in the
accretion disk of the central black hole or in the jets. Depending on the site
of acceleration, either the thermal photons of the accretion disk or the
synchrotron radiation in the jet may serve as the target needed for pion
production (e.g. \citet{NMB93,SS96}).

It should be noted that the detection of neutrinos emitted by AGNs would be a
smoking gun for hadron acceleration.

\subsubsection*{Galaxy clusters:}
The ionized gas in a galaxy cluster might act as the target for cosmic rays
escaping from cluster galaxies \citep{CB98}.

\subsubsection*{GRBs:}
Protons might be accelerated in gamma-ray bursts (GRBs) (e.g. \citet{WB97,WB00,Pao01}). The main advantage of
gamma-ray bursts is that they allow a precise timing, which reduces the
background considerably. In addition, assuming that two neutrinos emitted at
the same moment in a GRB located at a distance $d$ arrive at
different times $t$ and $t+\Delta t$, one may compute the speed difference
$\Delta v_{\nu}$ of the two neutrinos by means of
\[\frac{\Delta v_{\nu}}{c}\approx10^{-18}\left(\frac{d}{10\rm
Gpc}\right)^{-1}\left(\frac{\Delta t}{1\rm s}\right).\]
Hence GRBs could offer a possibility for checking neutrino dispersion.\vspace{2ex}

\begin{figure}
\includegraphics[width=8.3cm]{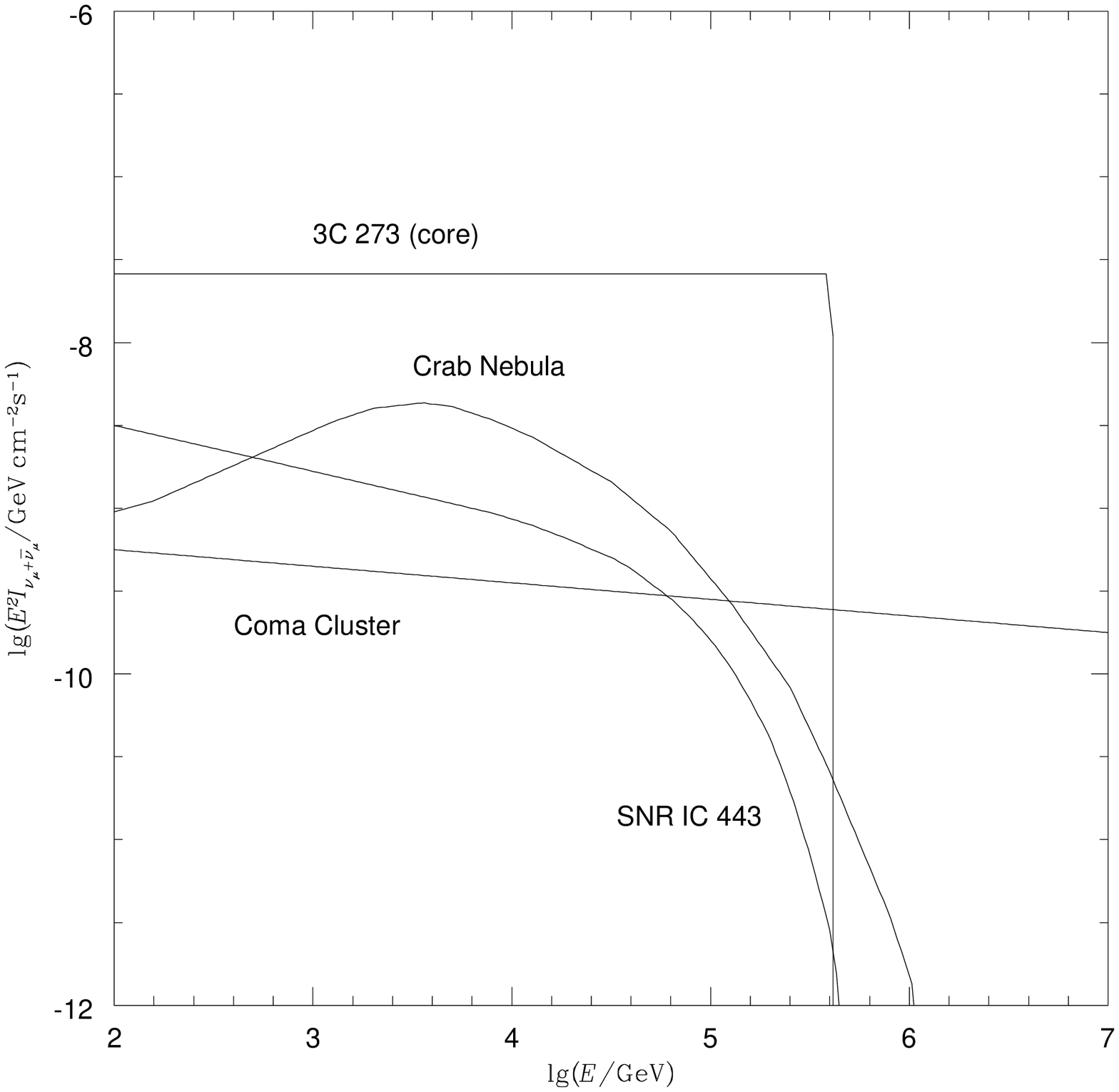}
\caption{Muon neutrino and antineutrino fluxes for 3C 273 \citep{NMB93,LM00}, the Crab
nebula \citep{BP97}, the supernova remnant SNR IC 443 \citep{GPS98,LM00}, and the Coma cluster
\citep{CB98}. No neutrino flavor oscillations are
assumed.}
\label{fig:fluxes}
\end{figure}

Fig.~\ref{fig:fluxes} shows the muon neutrino fluxes predicted for various
source models, if no neutrino flavor oscillations are assumed.
\begin{figure}
\includegraphics[width=8.3cm]{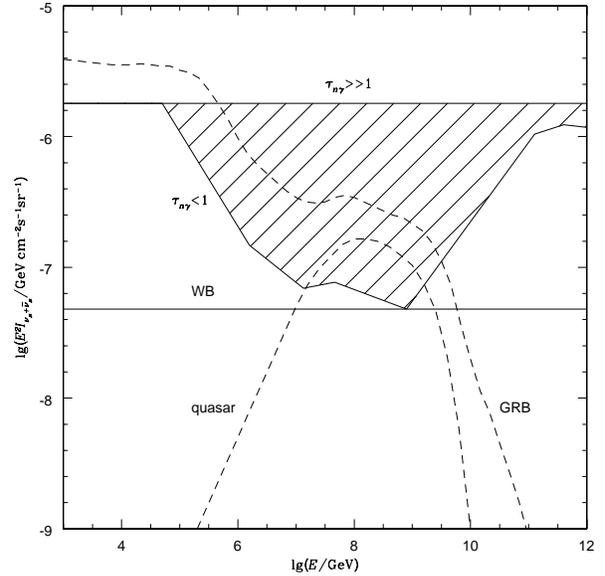}
\caption{Generic muon neutrino and antineutrino flux bound for sources with low
\emph{($\tau_{n\gamma}<1$)} and high neutron opacity
\emph{($\tau_{n\gamma}\gg1$)} (MPR flux bounds), and generic bound for neutrinos emitted by
quasars \emph{(quasar)}, as obtained by \citet{MPR01}. The shaded region is
the allowed range for the MPR flux bounds. Also shown are the
generic GRB flux bound given by \citet{Man01} \emph{(GRB)} and the
flux bound obtained by \citet{WB99} \emph{(WB)}. No neutrino flavor
oscillations are assumed.}
\label{fig:bounds}
\end{figure}

\section{Neutrino flux bounds}\label{sec:bounds}
The multitude (and uncertainty) of neutrino source models suggests the question
whether one can obtain a generic upper neutrino flux bound, thus
constraining the number of events to be expected from future neutrino
telescopes. 
In order to construct such a bound, we start by assuming that in a
neutrino source all protons are confined within the acceleration region (while
neutrons and neutrinos may escape), that there is a power law target photon
spectrum with index $\alpha$, $n_{\gamma}(E_{\gamma})\propto
E_{\gamma}^{-\alpha}$, and that the upper limit on the cosmic ray proton
spectrum as observed on Earth has the form $n_{p,\rm obs}(E_{p})\propto
E_{p}^{-2.75}$ between $3\times10^{6}$ and $10^{12}\ \rm GeV$ \citep{MPR01}.

Particle physics considerations then show that the neutrino production rate
$Q_{\nu_{\mu}}$ within the source is given by
\begin{equation}
Q_{\nu_{\mu}}(E_{\nu_{\mu}})\approx\left\{\begin{array}{ll}
  83.3Q_{n}(25E_{\nu_{\mu}}) & \mbox{\hspace*{3ex}(if $\alpha=1$)}\\
  416Q_{n}(25E_{\nu_{\mu}}) & \mbox{\hspace*{3ex}(if $\alpha=0$)}
\end{array}\right.,\label{eq:Qnu}
\end{equation}
where $Q_{n}$ denotes the neutron production rate \citep{MPR01}. The source contributes to
the cosmic ray proton flux, if the neutron escape probability $P_{{\rm
esc,}n}$ does not vanish. Indeed, the cosmic ray production rate can be
estimated as \citep{MPR01}
\begin{eqnarray}
Q_{\rm cr}(E_{p}) & \approx & Q_{n}(E_{p})\times P_{{\rm
esc,}n}(E_{p})\approx\frac{1-e^{-\tau}}{\tau}Q_{n}(E_{p})\nonumber\\
& \approx & Q_{n}(E_{p})\times\left\{\begin{array}{ll}
  1 & \mbox{\hspace*{3ex}($\tau<1$)}\\
  \tau^{-1} & \mbox{\hspace*{3ex}($\tau>1$)}
\end{array}\right..\label{eq:Qcr}\end{eqnarray}
Here $\tau$ constitutes the neutron optical depth for the source
radius. Finally, we note that the bolometric photon and neutrino luminosities
are related by $L_{\gamma}=2L_{\nu}$ for $\alpha=1$ and $L_{\gamma}\approx
L_{\nu}$ for $\alpha=0$ \citep{RM98}.

We can now give a formal recipe for constructing generic upper flux bounds \citep{MPR01}:
\begin{enumerate}
\item Get the neutron production rate $Q_{n}$. Note that due to the rescaling
in step 4, $Q_{n}$ need to be known up to a factor of proportionality only.
\item By means of Eqs.~\ref{eq:Qnu}
and~\ref{eq:Qcr} compute the corresponding neutrino and cosmic ray proton production
rates (i.e. $Q_{\nu_{\mu}}$ and $Q_{\rm cr}$).
\item Integrate the neutrino and cosmic ray proton production rates over all
sources in the universe to obtain the neutrino and cosmic ray proton intensities
$I$ at the Earth. Formally, this may be done using the formula
\begin{eqnarray*}
I(E)\propto\frac{1}{4\pi}\int_{z_{\rm min}}^{z_{\rm
max}}M(E,z)\frac{(1+z)^{2}}{4\pi d_{L}^{2}}\frac{\dd V_{\rm c}}{\dd
z}\frac{\dd P_{\rm source}}{\dd V_{\rm c}}\cdot\\
& & \mbox{\hspace*{-23.5ex}}\cdot Q((1+z)E,z)\dd z,
\end{eqnarray*}
where $z$ denotes the redshift, $d_{L}$ the luminosity distance, $V_{\rm c}$ the comoving
volume, and $\dd P_{\rm source}/\dd V_{\rm c}$ the source density, which
analogously to $Q_{n}$ in step 1 needs
to be known only up to a factor of proportionality. The
modification factor $M$ takes into account the attenuation processes along the
line of flight. For neutrinos, $M(E,z)=1$. The factor $1+z$ in the argument of
$Q$ describes the energy degradation due to the expansion of the
universe.
\item Normalize $I_{\rm cr}$ and thus $I_{\nu_{\mu}}$ so that $I_{\rm cr}$ is tangent to the observational upper
limit of the cosmic ray proton intensity or so that $\gamma$-rays aren't
overproduced (whichever is the more stringent condition).
\end{enumerate}
It should be noted that the bounds thus computed pertain to hadronic source
models only, so that there might be an additional contribution from
non-hadronic sources. One may now use this recipe for computing various
bounds:

Generic bounds \citep{MPR01}
are obtained, if one assumes that the neutron production spectrum is given by
$Q_{n}(E_{n})\propto E_{n}^{-1}\exp(-E_{n}/E_{\rm max})$ with a cutoff energy
$E_{\rm max}$ and that the source distribution follows that of galaxies and
AGNs. The corresponding cosmic ray proton flux as a function of $E_{\rm max}$ is
fitted to the observational bound. We shall refer to these bounds as the
Mannheim-Protheroe-Rachen (MPR) flux bounds.

In order to get a bound for EGRET blazars \citep{MPR01}, we assume that the production
rate of accelerated protons is of the form
$Q_{p}(E_{p})\propto E_{p}^{-2}\exp(-E_{p}/E_{\rm max})$, corresponding to
non-relativistic Fermi acceleration, and that the spectral index of the
target photon spectrum has the value $\alpha=1$. From $Q_{p}$ and the proton loss
time scale one may compute the proton spectrum $N_{p}$ in the source, and from
$N_{p}$ and the timescale of $p\gamma\longrightarrow n$ the neutron production
rate $Q_{n}$, which may serve as the starting point for the general recipe
outlined above. Here, as the production rates depend on the luminosity,
the average over the luminosity function $\dd N/\dd L$ is used:
\[\langle Q_{\nu_{\mu},\rm cr}(E)\rangle=\frac{\int Q_{\nu_{\mu},\rm
cr}(E,L)\frac{\dd N}{\dd L}\dd L}{\int\frac{\dd N}{\dd L}\dd L}.\]

For GRBs \citep{Man01} one takes the spectrum of accelerated protons to be of the form
$Q_{p}(E_{p})\propto E_{p}^{-2}$, but assumes that the photon target spectral
index is given by $\alpha=0$. The calculation can then be carried out
analogously to that for the blazar bound.

Finally, if it is assumed that the proton input spectrum is of the form
$Q_{p}(E_{p})\propto E_{p}^{-2}$ and that the sources are optically thin for
neutrons, one obtains the Waxman\,\&\,Bahcall bound \citep{WB99}.

In Fig.~\ref{fig:bounds} various upper flux bounds for the muon neutrino flux
from hadronic sources are shown.

\section{Event rates in neutrino telescopes}
\begin{figure*}
\includegraphics[width=8.3cm]{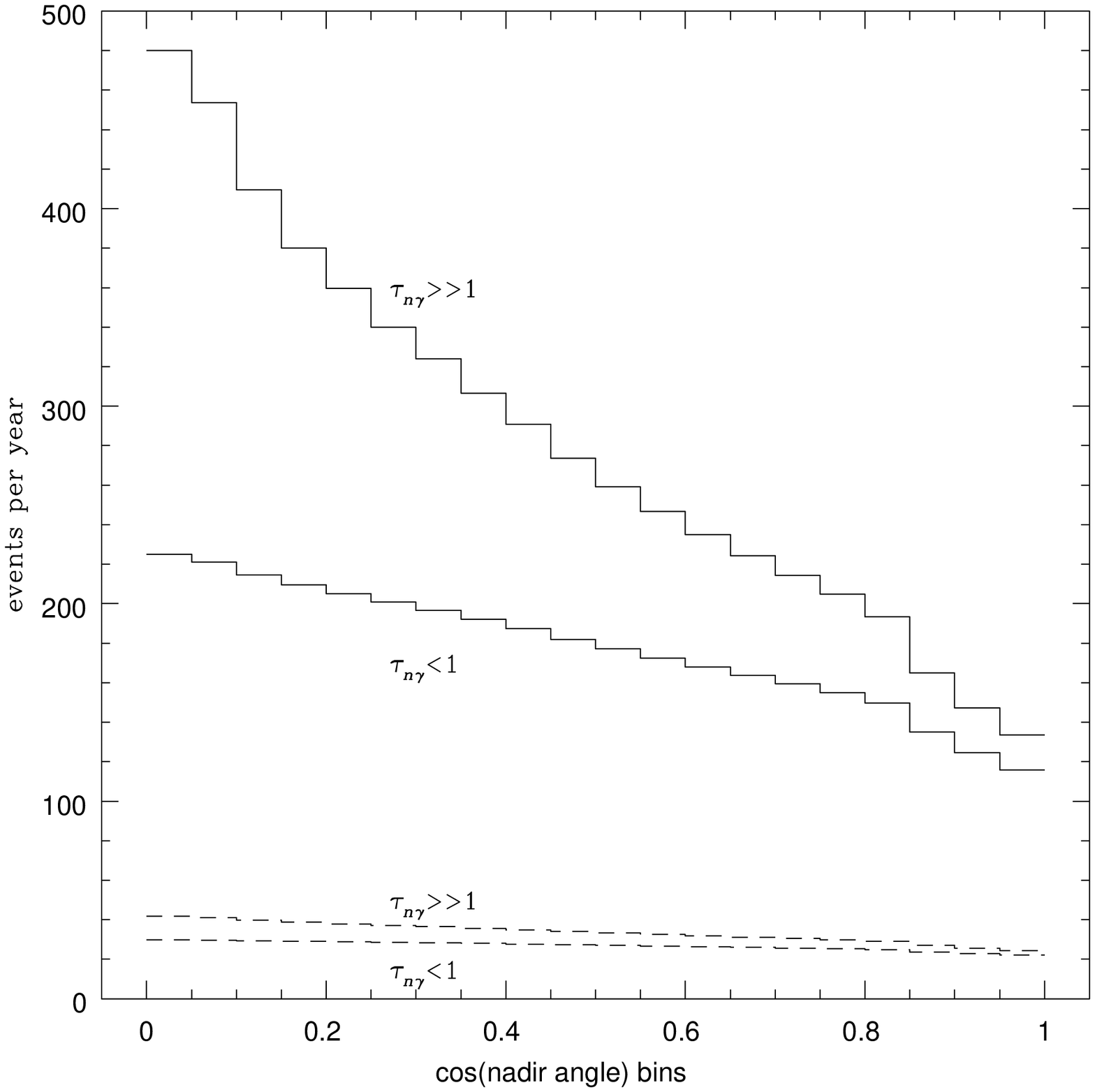}\hfill\includegraphics[width=8.3cm]{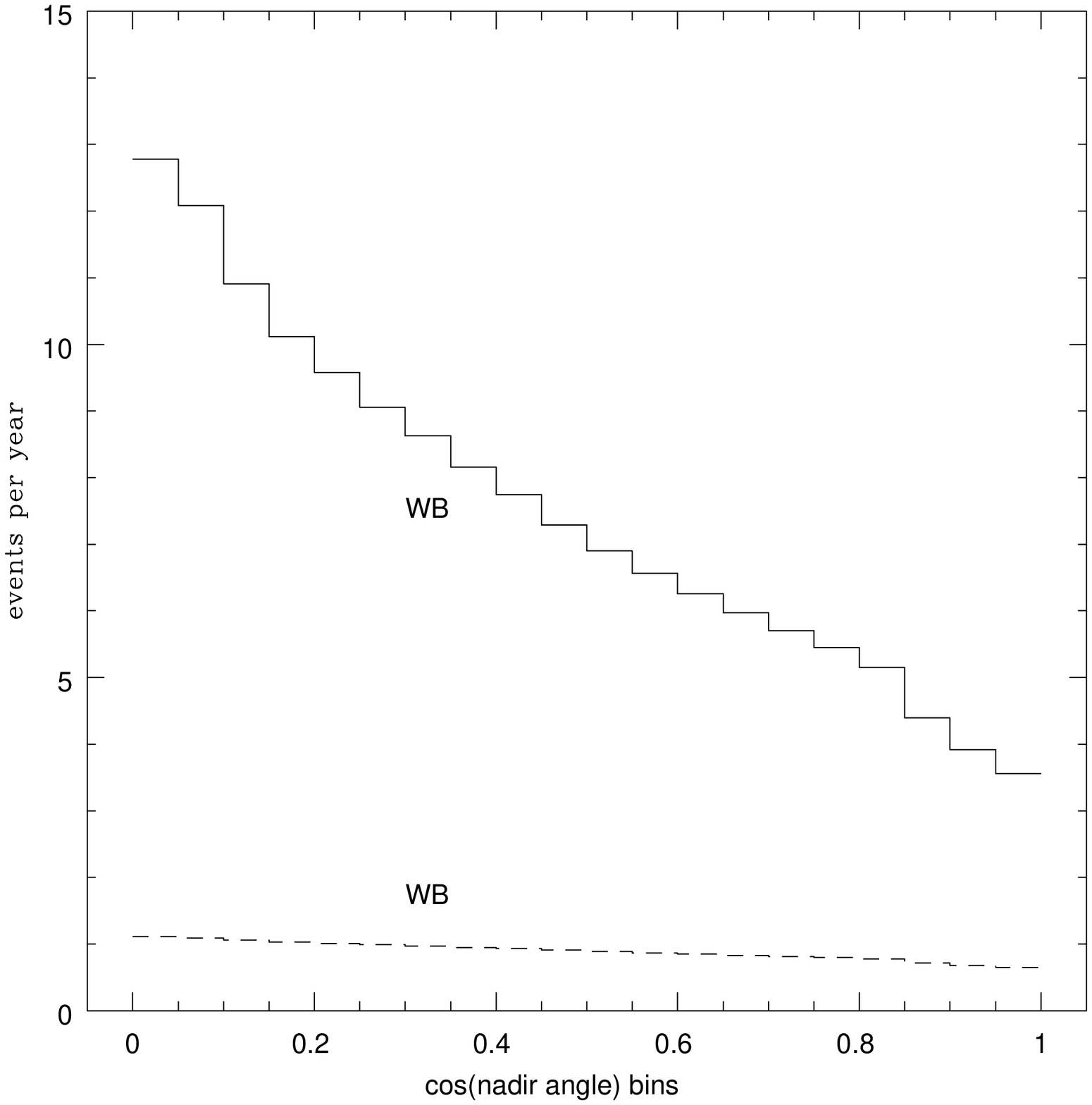}
\caption{Event rates of muon \emph{(solid)} and tauon neutrinos \emph{(dashed
lines)} for bins of the cosine of the nadir angle. \emph{Left:} Event rates
for the MPR flux bounds with high \emph{($\tau_{n\gamma}\gg1$)} and low neutron
source opacity \emph{($\tau_{n\gamma}<1$)}. \emph{Right:} Event rates for
the Waxman\,\&\,Bahcall flux bound.}
\label{fig:boundevents}
\end{figure*}
The decay of a charged pion leads to the production of one electron and two
muon neutrinos, where here and in the following no distinction between
particles and antiparticles is made. This implies that initially the ratio of
the various neutrinos should be given by
$Q_{\nu_{e}}:Q_{\nu_{\mu}}:Q_{\nu_{\tau}}=1:2:0$.

However, during the flight to Earth neutrino flavor oscillations should occur,
leading to an equipartition among the various flavors \citep{LP95,Hus00}. Hence the flux ratio as
measured on Earth should have the value
\[I_{\nu_{e}}:I_{\nu_{\mu}}:I_{\nu_{\tau}}=1:1:1,\]
so that a considerable amount of tauon neutrinos is to be expected, although
virtually no tauon neutrinos are produced within neutrino sources. Neutrinos
reaching a detector from below have to cross the inner Earth first. This
propagation can formally be described by means of the set \citep{HM01}
\begin{equation}
\frac{\dd I_{\nu_i}}{\dd t}=-\sigma_{i,\rm
tot}I_{\nu_{i}}+\sum_{k}\int_{E}^{\infty}\frac{\dd\sigma_{k\rightarrow
i}(E^{\prime},E)}{\dd E}I_{\nu_{k}}(E^{\prime}),
\label{eq:propagation}
\end{equation}
of cascade equations, where $\sigma_{i}$ denotes the total neutrino cross section of flavor $i$, the
sum runs over all flavors, and $\dd\sigma_{k\rightarrow i}/\dd E$ covers
both neutrino-nucleon and leptonic neutrino interactions, and the decay
of tauons. For the parton distributions we use the CTEQ5DIS functions
\citep{Lai00}, and we discretize the
transport equations. The resulting set of equations can then be solved
semi-analytically.

As a rough approximation to the event rate $\dot{N}$ due to neutrino detection
in water-based \v{C}erenkov telescopes, we employ the formula
\begin{eqnarray}
\lefteqn{\dot{N}_{i}(E)\approx\frac{\rho A_{\rm
eff}}{m_{p}}\left(\int_{E}^{\infty}\dd E^{\prime}\frac{\dd\sigma_{i}^{\rm CC}}{\dd
E^{\prime}}I_{i}(E^{\prime})L_{\rm det}\right.}\nonumber\\
& & \hspace*{-5ex}\mbox{}+\left.\int_{L}\dd
x\int_{E}^{\infty}\dd E^{\prime}\frac{\dd\sigma_{i}^{\rm CC}}{\dd
E^{\prime}}(E_{0}^{(i)}(E^{\prime},x))I_{i}(E_{0}^{(i)}(E^{\prime},x))\right).
\label{eq:detection}
\end{eqnarray}
Here again $i$ denotes the flavor. $A_{\rm eff}$ constitutes the
effective detector area, $L$ the detector height, $\rho$ the matter
density, and $m_{p}$ the proton mass. $E_{0}^{(i)}(E^{\prime},x)$ is the energy (of the
lepton $i$) which is degraded to $E^{\prime}$ along a path of length
$x$. In the following, we shall assume $A_{\rm eff}=1\ {\rm km}^{2}$, $L=1\ {\rm km}$, and $\rho=1\ {\rm g}/{\rm cm}^{3}$.

Whereas the first term on the right hand side of Eq.~\ref{eq:detection} describes fully
contained events, the second term covers leptons originating outside the
detector. It may be neglected for electrons and tauons.

Finally, it ought to be noted that the decay of tauons results in the
production of both electrons and muons and hence contributes to the respective
event rates.

\section{Results}
The formalism outlined in the previous section can be used to obtain the event
rates corresponding to various of the flux bounds discussed in
Sect.~\ref{sec:bounds}. This has been carried out for the Waxman\,\&\,Bahcall and
the MPR flux bounds. The results (for a lower energy cutoff of 100 GeV) are given in Fig.~\ref{fig:boundevents}.

As in Fig.~\ref{fig:boundevents} different bins correspond to the same solid
angle, the form of the curves is a direct measure of attenuation effects due
to crossing the Earth. This suggests that one might use the event rate as a
function of the nadir angle in order to perform a tomography of the inner
Earth \citep{JRF99}.

To this end, we start by noting that the column density $X$ may be considered
as the one-dimensional Radon transform of the number density $\rho$:
\begin{equation}
X(p,\bvec{\xi})=\int\rho(\bvec{r})\delta(p-\bvec{\xi}\bvec{r})\dd^{2}r.
\label{eq:radon1d}
\end{equation}
Here, $p$ and $\bvec{\xi}$ are the distance from the middle of the Earth and the normal
vector of the line of integration, respectively. Now, using the inversion formula for two-dimensional Radon transforms \citep{Dea93} one
can show that $X(p,\bvec{\xi})\equiv X(p)$ and that one may thus invert Eq.~\ref{eq:radon1d} by means of
\begin{equation}
\rho(r)=-\frac{1}{2\pi r}\frac{\dd}{\dd
r}\int_{-\sqrt{R^{2}-r^{2}}}^{+\sqrt{R^{2}-r^{2}}}X(\sqrt{r^{2}+q^{2}})\dd q,
\label{eq:inversion}
\end{equation}
where $\rho(r)\equiv0$ for $r>R$. However, if the initial neutrino spectrum (before crossing the Earth) is
assumed to be known, one may deduce the column density directly from the
observed event rate. Then Eq.~\ref{eq:inversion} yields the density, so
that a tomography is indeed possible.

Unfortunately, there is an important caveat: Due to the statistical error $\sigma$ in
the detection rates, the column densities can be known only with a rather
limited precision. We take this into account by either adding or subtracting
$\sigma$ throughout and comparing the results thus obtained with that for
$\sigma=0$.

\begin{figure*}
\includegraphics[width=8.3cm]{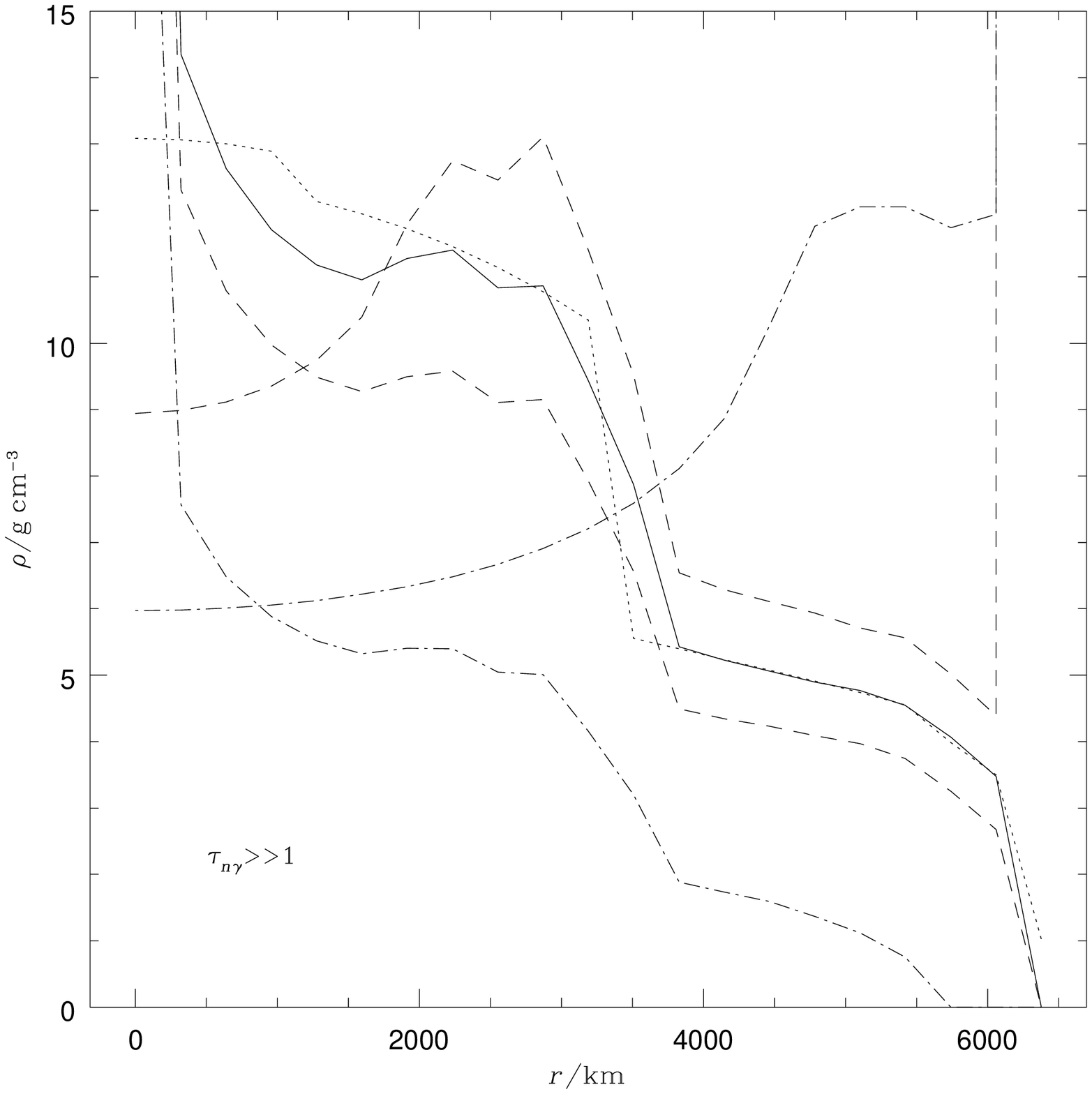}\hfill\includegraphics[width=8.3cm]{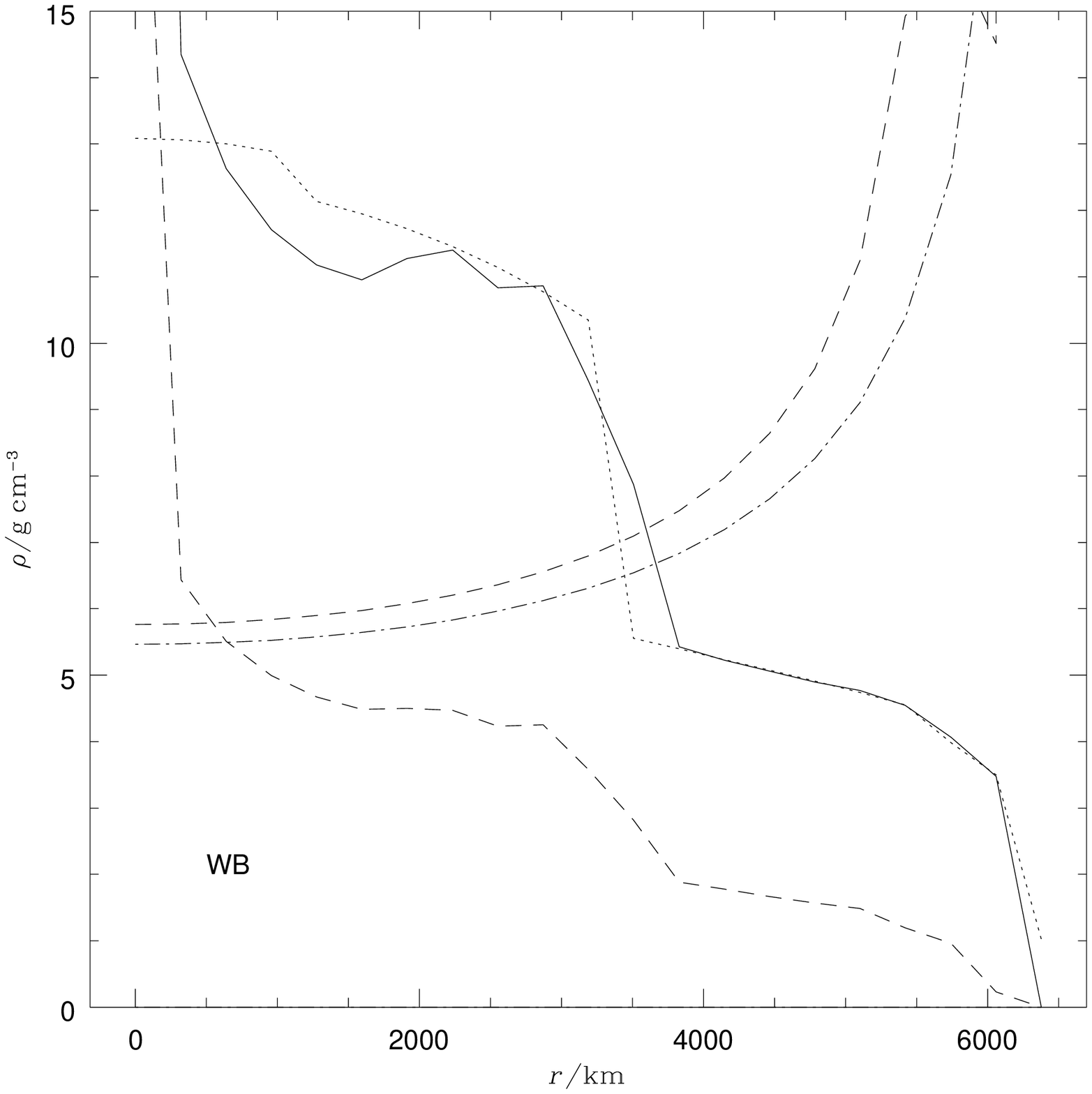}
\caption{\emph{Left:} Earth density $\rho$ as a function of the radius $r$ obtained by
performing a tomography on the event rates $\dot{N}$ of
Fig.~\ref{fig:boundevents} for the MPR flux bound with high neutron source opacity. Shown are the results for no statistical error
\emph{(solid)}, for $\dot{N}_{\mu}\pm\sqrt{\dot{N}_{\mu}}$ \emph{(dashed)},
and for $\dot{N}_{\tau}\pm\sqrt{\dot{N}_{\tau}}$ \emph{(dot-dashed
lines)}. A lower energy cutoff of 100 GeV is assumed. For comparison the
``correct'' density according to \citet{DA81} is included \emph{(dotted line)}. \emph{Right:} Same as
left figure, but for the Waxman\,\&\,Bahcall flux bound.}
\label{fig:tomography}
\end{figure*}

The results of this analysis for a one-year run are shown in Fig.~\ref{fig:tomography} for an
MPR and for the Waxman\,\&\,Bahcall flux bound. One can see that the maximum event
rates to be expected for next-generation neutrino telescopes put severe constraints on a
tomography of the inner Earth.

\begin{acknowledgements}
This work has been supported by the Studienstiftung des deutschen Volkes.
\end{acknowledgements}
\end{document}